\documentclass[a4paper,12pt]{article}
\usepackage{amsmath,amsthm,amssymb,bbm}

\usepackage[english]{babel}

\title{\bf Entangling two unequal atoms\\ through a common bath}

\author{F. Benatti$^{a,b}$, 
R. Floreanini$^{b}$ and U. Marzolino$^{a,b}$\\
\small ${}^a$Dipartimento di Fisica Teorica, Universit\`a di Trieste, 
34014 Trieste, Italy\\
\small ${}^b$Istituto Nazionale di Fisica Nucleare, Sezione di Trieste,
34014 Trieste, Italy}

\date{\null}

\begin{document}

\maketitle

\begin{abstract}
\noindent
The evolution of two, non-interacting two-level atoms immersed in a weakly coupled
bath can be described by a refined, time coarse grained Markovian evolution,
still preserving complete positivity. We find that this improved reduced dynamics
is able to entangle the two atoms even when their internal frequencies are unequal,
an effect which appears impossible in the standard {\it weak coupling limit} approach. 
We study in detail this phenomenon for an environment made of quantum fields.
\end{abstract}

\section{Introduction}

Independent, non-interacting atoms immersed in a common bath, 
represent an example of an open quantum system, {\it i.e.} a subsystem
in interaction with a large environment \cite{Gorini}-\cite{B}.
Their time evolution can be given as a reduced dynamics, obtained by eliminating
the degrees of freedom of the environment and by subsequently performing a
Markovian, memoryless approximation, justified by the very rapid
decay of correlations in the bath. The resulting atom evolution
is irreversible and incorporates dissipative and noisy effects induced by the environment.

In many physical instancies, the atoms can be treated in a non-relativistic approximation,
as independent finite-level systems, with negligibly small size. On the other hand, the environment
can be modeled by a set of weakly coupled quantum fields, typically the electromagnetic field,
in a given temperature state, interacting with the atoms through a dipole type coupling \cite{Milonni}. 
Although ignoring the internal atom dynamics 
and the full vectorial structure of the electromagnetic field,
this simplified setting is nevertheless perfectly adequate for studying 
the behaviour of physical systems
like ions in traps, atoms in optical cavities and fibers, 
impurities in phonon fields \cite{Milonni}-\cite{Puri}.

The derivation of an acceptable subdynamics for the atoms
is notoriously tricky \cite{Dumcke}, \cite{Gorini}-\cite{Alicki}
and time evolutions that are not even positive have been adopted in the literature
in order to describe their physical properties \cite{BF,ABF,BF2}.
Instead, a physically consistent time evolution
for the atom subsystem can be obtained through a suitable coarse grained
procedure, within the    
weak coupling approximation \cite{Gorini,Alicki,Alicki1,Lidar,Schaller}: the resulting subdynamics
is described by a one parameter ($\equiv$ time) family of completely
positive maps that form a quantum dynamical semigroup.

We shall explicitly discuss below such a derivation
for a subsystem composed by two, unequal, mutually non-interacting atoms.
For simplicity we shall restrict the
attention to two-level atoms weakly coupled to a collection
of independent, free, massless scalar fields in $3+1$ space-time
dimensions, assumed to be in a state at temperature $T\equiv1/\beta$.

As well known, the interaction with an environment
usually leads to decoherence and noise, typical mixing enhancing
phenomena. Therefore, one generally expects that quantum correlations
that might have been present at the beginning between the
two atoms be destroyed when they are immersed in the bath.

However, an external environment can also mediate indirect
interactions between otherwise totally decoupled subsystems
and therefore a mean to correlate them. This phenomenon is 
generic on a short, microscopic time-scale where the dynamics is unitary
and reversible; however, its persistence is
not expected in general on
longer time-scales where irreversible, dissipative and
decohering effects described by Markovian master equations appear.

Nevertheless, there are instances where purely dissipative, non-Hamiltonian
contributions to the master equation can lead to entanglement generation
\cite{Plenio}-\cite{Paz},\cite{Ficek}. This is due to the particular
form of the (Kraus) operators appearing there, which couple indirectly ({\it i.e.}
not dynamically) the two subsystems.
This phenomenon has been established in the case of subsystems formed by two
identical two-level systems \cite{BFP}-\cite{BF5} 
or harmonic oscillators \cite{BF6,BFR} evolving with 
a reduced dynamics obtained via the so-called {\it weak coupling limit} \cite{Davies}.
This technique is applicable when the time-scale over which the
dissipative effects become visible is so large that the free dynamics of
the subsystems can be effectively averaged out, thus eliminating
very rapid oscillations; this typically occurs for environments
with very fast decaying correlations \cite{Spohn,Alicki,BF}.
Nevertheless, it turns out that, in the case of atoms with unequal frequencies,
this procedure of averaging out fast oscillations prevents
generation of entanglement; in other terms, the environments for
which the {\it weak coupling limit} procedure is justified are unable
to correlate two atoms with different internal frequencies,
while they do when the atoms are identical.

In the following, we shall study in detail the conditions
that allow the two otherwise independent, unequal atoms to become
initially entangled through the action of the environment, 
when the {\it weak coupling limit} procedure is not applicable.%
\footnote{A preliminary investigation on these topics in the case of
two qubits weakly coupled to a Ohmic bath has been reported in \cite{BFM}.}
Following \cite{Schaller}, we shall instead derive a time evolution for the two atoms
that allows a finite coarse-graining time interval: remarkably, it results still
expressible in terms of a completely positive quantum dynamical semigroup, which
reduces to the standard one obtained through the {\it weak coupling limit} in the
limit of an infinitely large coarse-graining time interval.
We shall see that in this refined framework the sharp dependence on the atom frequencies 
of the entanglement capability of the environment mentioned above looks more like
a mathematical artifact than a real physical effect.

\section{Two-atom reduced dynamics}

As explained above, we shall study the behaviour of a system composed by two, unequal
two-level atoms, that start interacting at time $t=\,0$ with an environment made of
a collection of independent, massless, 
scalar quantum fields at temperature $1/\beta$.%
\footnote{As we shall see in the following, the choice of modelling the environment in terms of
relativistic quantum fields (at finite temperature) allows an analytic treatment of
the reduced two-atom dynamics, without additional approximations
besides the Born and Markov ones (see below).}
We are interested in the evolution of the atoms as open quantum systems and not
in the details of their internal dynamics; therefore, we shall model them,
in a nonrelativistic way, as simple qubits,
described in terms of a two-dimensional
Hilbert space.

In absence of any interaction with the external fields,
the single atom internal dynamics can thus be taken to be driven by a generic
$2\times 2$ hamiltonian matrix. As a result, the total atom Hamiltonian $H_S$
can be expressed as:
\begin{equation}
H_S=H_S^{(1)}+H_S^{(2)}\ ,
\quad
H_S^{(\alpha)}={\omega_\alpha\over 2}\, \vec n\cdot\vec\sigma^{(\alpha)}
\equiv{\omega_\alpha\over 2}\sum_{i=1}^3 n_i\, \sigma_i^{(\alpha)}
\ ,\quad \alpha=1,2\ ,
\label{1}
\end{equation}
where $\sigma_i^{(1)}=\sigma_i\otimes{\bf 1}$ and
$\sigma_i^{(2)}={\bf 1}\otimes\sigma_i$ are the basis
operators pertaining to the two different atoms,
with $\sigma_i$, $i=1,2,3$ the Pauli matrices, $n_i$
the components of a unit vector, while $\omega_\alpha$ represent the
gaps between the two energy eigenvalues of the two atoms.

As mentioned in the introductory remarks, the coupling of the atoms 
with the external fields is assumed to be weak, so that
the dipole approximation results appropriate \cite{Milonni}. In our simplifying settings that
ignore spinorial indices, the interaction term can then
be described by an hamiltonian $H'$ that is linear in both atom and field
variables:
\begin{equation}
H'=\sum_{i=1}^3\Big(\sigma_i^{(1)}\otimes \Phi_i[f^{(1)}]
+\sigma_i^{(2)}\otimes \Phi_i[f^{(2)}]\Big)\ .
\label{2}
\end{equation}
The operators
$\Phi_i(t,\vec x)$ represent the set of external quantum fields,
taken to be spinless and massless for simplicity. 
They evolve in time as free relativistic fields with a standard
Hamiltonian $H_\Phi$ \cite{Haag}.
The atoms are assumed to have a spatial extension described by the two
functions $f^{(\alpha)}(\vec x)$, $\alpha=1,2$. To be more specific, we shall choose 
for the atoms a common profile $f(\vec x)$ of
spherically symmetric shape, with size $\varepsilon$:
\begin{equation}
f(\vec x)={1\over\pi^2}{(\varepsilon/2)\over \big[|\vec x|^2 
+(\varepsilon/2)^2\big]^2}\ ,
\label{3}
\end{equation}
and position the first atom at the origin of the reference frame,
so that $f^{(1)}(\vec x)\equiv f(\vec x)$, while the second is
displaced by an amount $\vec\ell$ with respect to it, 
$f^{(2)}(\vec x)= f(\vec x+\vec\ell\,)$. Since the atom-field interaction 
takes place on the whole region occupied by the atoms, 
the field operators entering the interaction Hamiltonian 
above are smeared over the atom spatial extension:
\begin{equation}
\Phi_i[f^{(\alpha)}]=\int d^3 x\, f^{(\alpha)}(\vec x\,)\, 
\Phi_i(0,\vec x\,)\ ,\quad \alpha=1,2\ .
\label{4}
\end{equation}

The total Hamiltonian $H$ describing the complete system, the two atoms
together with the external fields $\Phi_i$, can thus be written as
\begin{equation}
H=H_S +H_\Phi +\lambda H'\equiv H_0 + \lambda H'\ ,
\label{5}
\end{equation}
with $\lambda$ a small coupling constant.%
\footnote{For simplicity, we have assumed that the environment couples with
the same strength to the two atoms: this makes the following analytic derivation of
the master equation more transparent, without compromising its generality.}
Through the standard Liouville-von Neumann equation,
$\partial_t\rho_{\rm tot}(t)=-i[H,\ \rho_{\rm tot}(t)]$,
it generates the evolution in time of the state of the total system,
described in general by a density matrix $\rho_{\rm tot}$,
starting at $t=\,0$ from the initial
configuration: $\rho_{\rm tot}(0)$.

We shall assume the atom and the fields to be initially prepared 
in an uncorrelated state, with the fields in
the temperature state $\rho_\beta$ and the atoms in a
generic initial state $\rho(0)$, so that
$\rho_{\rm tot}(0)=\rho(0)\otimes \rho_\beta$.
The reduced time evolution of the two atoms is then obtained
by integrating over the unobserved field degrees of freedom
and is formally given by the transformation map:
$\rho(0)\mapsto\rho(t)\equiv{\rm Tr}_\Phi[\rho_{\rm tot}(t)]$.
This map is in general very complicated, because of nonlinearities and memory effects; 
nevertheless, it can be approximated by a linear, memoryless map when the coupling
with the environment is small 
and its own internal dynamics is sufficiently fast \cite{Gorini}-\cite{BF}. 
Indeed, in such cases the details of the internal environment dynamics result 
irrelevant, being the time scale of the subsystem evolution
typically very long compared with the decay time
of the correlations in the bath.%
\footnote{A discussion on the validity of this so-called Markovian approximation
is reported in \cite{Alicki1}. There, a non-Markovian weak coupling approximation
of the reduced dynamics is also introduced; it leads to a two-parameter family
of dynamical maps, with a time-dependent generator \cite{Alicki}.
We stress that this approach is completely different from the one discussed below,
which instead describes the reduced two-atom dynamics in terms of a Markovian, 
one parameter semigroup. In particular, while in \cite{Alicki1} the standard {\it weak coupling
limit} can be reached only in the asymptotic, long-time regime,
in the treatment presented below it can always be obtained for any time
by letting the coarse-graining parameter become large.}

In order to derive the equation obeyed by the reduced density matrix
$\rho(t)$ in the case at hand, it is convenient to work in the interaction
representation
\begin{equation}
\tilde\rho_{\rm tot}(t)=e^{itH_0}\ \rho_{\rm tot}(t)\ e^{-itH_0}\ ,
\label{6}
\end{equation}
so that
\begin{equation}
{\partial\tilde\rho_{\rm tot}(t)\over \partial t}=-i \lambda\Big[H'(t),\, \tilde\rho_{\rm tot}(t)\Big]\ ,
\qquad H'(t)=e^{itH_0}\ H'\ e^{-itH_0}\ .
\label{7}
\end{equation}
One then focuses on the changes of the reduced state 
$\tilde\rho(t)\equiv{\rm Tr}_\Phi[\tilde\rho_{\rm tot}(t)]$ 
over a time interval $\Delta t$; by taking the trace over the field variables
of the integrated version of the equation (\ref{7}) one gets (to lowest order
in $\lambda$):
\begin{eqnarray}
\nonumber
{\tilde\rho(t+\Delta t)-\tilde\rho(t)\over \Delta t}&=&
{1\over\Delta t}\int_t^{t+\Delta t} ds\ {\partial\tilde\rho(s)\over \partial s}\\
&=&-{\lambda^2\over \Delta t}\int_t^{t+\Delta t} dt_1
\int_t^{t_1} dt_2\ {\rm Tr}_\Phi\Big( \big[ H'(t_1),\big[H'(t_2),\, \tilde\rho_{\rm tot}(t)\big]\big]\Big)
+O(\lambda^4)\ .\ \
\label{8}
\end{eqnarray}
One notices that the variation of $\tilde\rho(t)$ starts to become relevant at order $\lambda^2$,
{\it i.e.} on time scales of order $\tau=\lambda^2 t$. Then, one can equivalently write:
\begin{equation}
{\tilde\rho(t+\Delta t)-\tilde\rho(t)\over \Delta t}=
{1\over\Delta t}\int_\tau^{\tau+\lambda^2\Delta t} ds\ {\partial\tilde\rho(s/\lambda^2)\over \partial s}\ ,
\label{9}
\end{equation}
so that in the limit of small $\lambda$ (and finite $\Delta t$)
one can readily approximate the r.h.s. of (\ref{9}) with $\partial_t\tilde\rho(t)$.
At this point, one further observes that the environment, containing an infinite number of 
degrees of freedom, is much larger than the subsytem immersed in it, so that its
dynamics is hardly affected by its presence. It is therefore justified to replace 
in the double integral of (\ref{8}) the evolved total state $\tilde\rho_{\rm tot}(t)$ with
the product state $\tilde\rho(t)\otimes\rho_\beta$, taking the initial 
state $\rho_\beta$ as a reference state for the bath \cite{Gorini}-\cite{Alicki}.

Returning to the Schr\"odinger representation, one finally gets the following linear,
Markovian master equation for the two-atom state $\rho(t)$:
\begin{equation}
{\partial\rho(t)\over \partial t}= -i \big[H_S,\, \rho\big]
 + {\cal D}[\rho(t)]\ ,
\label{10}
\end{equation}
where the bath-dependent contribution ${\cal D}[\rho(t)]$ contains both 
an Hamiltonian and a dissipative term
\begin{equation}
{\cal D}[\rho(t)]=-i\big[H_{12},\rho(t)\big]+{\cal L}[\rho(t)]\ ,
\label{11}
\end{equation}
with
\begin{eqnarray}
\label{12}
&&H_{12}={i\lambda^2\over2\Delta t}\int_0^{\Delta t}ds_1\int_0^{\Delta t} ds_2\
\theta(s_1-s_2)\ {\rm Tr}_\Phi\Big(\rho_\beta\, \big[ H'(s_1),\, H'(s_2)\big]\Big)\ ,\\
\label{13}
&&{\cal L}[\rho(t)]={\lambda^2\over \Delta t} {\rm Tr}_\Phi\Bigg[
L\, \big(\rho(t)\otimes\rho_\beta\big)\,L-{1\over2}\Big\{ L^2,\rho(t)\otimes\rho_\beta\Big\}\Bigg]\ ,
\quad L=\int_0^{\Delta t}ds\, H'(s)\ ,\qquad
\end{eqnarray}
the curly brackets representing the anticommutator, while $\theta(s)$ is the step function.

It is important to observe that, for any interval $\Delta t$, 
the master equation (\ref{10})-(\ref{13}) generates
a quantum dynamical semigroup of completely
positive maps. Indeed, the generator in the r.h.s. of (\ref{11}), besides the
Hamiltonian piece, contains a dissipative term which turns out to be itself completely positive,
being the composition of two completely positive maps, the trace over the environment degrees
of freedom and a linear operator on the total system, 
written in canonical Stinespring form \cite{Alicki2,Takesaki,B}.
Notice that, on the contrary, in the usual {\it weak coupling limit} approach 
to the derivation of a Markovian master equation,
complete positivity is ensured by an ergodic average prescription, that, as mentioned in the
Introduction, eliminates fast oscillating terms \cite{Dumcke,Davies}; 
in the present formalism, this corresponds
to letting the time coarse-graining parameter $\Delta t$ going to infinity: it is thus applicable
only to environments with sharp decaying correlations. In the following, we shall instead keep
$\Delta t$ finite and consider therefore more general situations.

\section{Master equation}

For the case at hand, a more explicit expression for the generator
in (\ref{11}) can be obtained by recalling (\ref{2}) and (\ref{4}).
Indeed, after straightforward manipulations, the master equation driving the
dissipative dynamics of the two atoms state takes the following 
Kossakowski-Lindblad form \cite{Gorini2,Lindblad}
\begin{equation}
{\partial\rho(t)\over \partial t}= -i \big[H_{\rm eff},\, \rho(t)\big]
 + {\cal L}[\rho(t)]\ ,
\label{14}
\end{equation}
with
\begin{equation}
H_{\rm eff}=H_S-\frac{i}{2}\sum_{\alpha,\beta=1}^2\sum_{i,j=1}^3
H_{ij}^{(\alpha\beta)}\ \sigma_i^{(\alpha)}\,\sigma_j^{(\beta)}\ ,
\label{15}
\end{equation}
and
\begin{equation}
{\cal L}[\rho]= \sum_{\alpha,\beta=1}^2\sum_{i,j=1}^3 C_{ij}^{(\alpha\beta)}\bigg[
\sigma_j^{(\beta)}\rho\,\sigma_i^{(\alpha)} 
-{1\over2}\Big\{\sigma_i^{(\alpha)}\sigma_j^{(\beta)},\ \rho\Big\}\bigg]\ .
\label{16}
\end{equation}
The coefficients of the Kossakowski matrix $C_{ij}^{(\alpha\beta)}$
and of the effective Hamiltonian $H_{\rm eff}$ are determined by
the field correlation functions in the thermal state $\rho_\beta$:
\begin{equation}
G_{ij}^{(\alpha\beta)}(t-t')=\int d^3x\, d^3y\, f^{(\alpha)}(\vec x)\,
f^{(\beta)}(\vec y)\ \langle \Phi_i(t,\vec x)\Phi_j(t',\vec y)\rangle\ ,
\label{17}
\end{equation}
through their Fourier,
\begin{equation}
{\cal G}_{ij}^{(\alpha\beta)}(z)=\int_{-\infty}^{\infty} dt \, e^{i z t}\, 
G_{ij}^{(\alpha\beta)}(t)\ ,
\label{18}
\end{equation}
and Hilbert transform,
\begin{equation}
{\cal K}_{ij}^{(\alpha\beta)}(z)=\int_{-\infty}^{\infty} dt \, {\rm sign}(t)\, 
e^{i z t}\, G_{ij}^{(\alpha\beta)}(t)=
\frac{P}{\pi i}\int_{-\infty}^{\infty} dw\ \frac{ {\cal G}_{ij}^{(\alpha\beta)}(w) }{w-z}
\ ,
\label{19}
\end{equation}
respectively ($P$ indicates principle value). 

More specifically, one finds that the Kossakowski matrix reads:
\begin{eqnarray}
\nonumber
C_{ij}^{(\alpha\beta)}&=&\lambda^2\!\sum_{\xi,\xi'=\{+,-,0\}} \sum_{k,l=1}^3\ 
e^{i(\xi\omega_\alpha+\xi'\omega_\beta)\Delta t/2}\ \psi_{ki}^{(\xi)}\, \psi_{lj}^{(\xi')}\\
&&\times{\Delta t\over 2\pi}\int_{-\infty}^\infty d\omega\ 
{\cal G}_{kl}^{(\alpha\beta)}(\omega)\, 
{\sin\big[(\xi\omega-\omega_\alpha)\Delta t/2\big]\over(\xi\omega-\omega_\alpha)\Delta t/2}\
{\sin\big[(\xi'\omega+\omega_\beta)\Delta t/2\big]\over(\xi'\omega+\omega_\beta)\Delta t/2}\ ,
\label{20}
\end{eqnarray}
where
\begin{equation}
\psi_{ij}^{(0)}=n_i\, n_j\ ,\qquad 
\psi_{ij}^{(\pm)}={1\over 2}\big(\delta_{ij} - n_i\, n_j\pm i\epsilon_{ijk} n_k\big)\ ,
\label{21}
\end{equation}
are the components of auxiliary three-dimensional tensors, giving the free evolution
of the atom operators:
\begin{equation}
\sigma_i^{(\alpha)}(t)=e^{itH_S} \sigma_i^{(\alpha)}\, e^{-itH_S}=
\sum_{\xi=\{+,-,0\}} \sum_{j=1}^3 e^{i\xi\omega_\alpha t}\
\psi_{ij}^{(\xi)}\ \sigma_j^{(\alpha)}\ .
\nonumber
\end{equation}
The $6\times 6$ matrix $C_{ij}^{(\alpha\beta)}$ turns out to be non-negative, 
since, as already mentioned,
the evolution generated by (\ref{9}) is completely positive.%
\footnote{On the other hand, let us remark
that direct use of the standard second order perturbative approximation
({\it e.g.} see \cite{Ficek,Puri}) often leads to physically inconsistent
results \cite{BF,ABF,BF2}, giving a finite time evolution for $\rho(t)$ that in general does 
not preserve the positivity of probabilities.}
An expression similar to the one in (\ref{20}) holds also for $H_{ij}^{(\alpha\beta)}$ in (\ref{15}),
with ${\cal G}_{kl}^{(\alpha\beta)}(\omega)$ replaced by 
${\cal K}_{kl}^{(\alpha\beta)}(\omega)$.

For simplicity, the fields giving rise to the environment are taken to be 
independent and further assumed to obey a free evolution; in this case, 
one finds:
\begin{equation}
\langle \Phi_i(x)\Phi_j(y)\rangle\equiv {\rm Tr}\big[\Phi_i(x)\Phi_j(y)\rho_\beta\big]
=\delta_{ij}\, G(x-y)\ ,
\label{22}
\end{equation}
where $G(x-y)$ is the standard four-dimensional Wightmann function for
a single relativistic scalar field in a state at inverse temperature $\beta$ \cite{Haag},
that, with the usual $i\varepsilon$ prescription, can be written as:
\begin{equation}
G(x)=\int \frac{d^4 k}{(2\pi)^{3}}\, \theta(k^0)\, \delta(k^2)
\Big[\big(1+{\cal N}(k^0)\big)\, e^{-ik\cdot x}+{\cal N}(k^0)\, e^{ik\cdot x}
\Big]e^{-\varepsilon k^0}\ ,
\label{23}
\end{equation}
where
\begin{equation}
{\cal N}(k^0)=\frac{1}{e^{\beta k^0} -1}\ .
\label{24}
\end{equation}
Although the $i\varepsilon$ prescription, assuring the convergence
of the integral in (\ref{23}), originates from causality requirements,
in the present setting it can be related to the finite size
of the two atoms. Indeed, the correlations in (\ref{17}) actually involve the Fourier transform
${\hat f}(\vec k)=\int d^3x\, e^{i\vec k\cdot\vec x}\, f(\vec x)$
of the shape function $f(\vec x)$ in (\ref{3}); it can be easily computed to be
${\hat f}(\vec k\,)=e^{-|\vec k\,| \varepsilon/2}$.
Inserting it back in (\ref{17}),
this contribution can be conveniently attached to the
definition of the Wightmann function $G(x)$, so that the integrand in (\ref{23})
gets an extra $e^{-\varepsilon k^0}$ overall factor. 

Using (\ref{23}) and (\ref{24}), the Fourier transform in (\ref{18})
can now be explicitly evaluated; taking for simplicity the limit of pointlike atoms,
(the size $\varepsilon$ can be taken to vanish since it does not play any more
the role of a regularization parameter), one gets:
\begin{equation}
{\cal G}_{ij}^{(\alpha\beta)}(\omega)=
\delta_{ij}\ {\cal G}^{(\alpha\beta)}(\omega)\ ,
\label{25}
\end{equation}
with:
\begin{eqnarray}
\nonumber
&&{\cal G}^{(11)}(\omega)={\cal G}^{(22)}(\omega)=\frac{1}{2\pi} \frac{\omega}{1-e^{-\beta \omega}}\ ,\\
\label{26}
&&{\cal G}^{(12)}(\omega)={\cal G}^{(21)}(\omega)=\frac{1}{2\pi} \frac{\omega}{1-e^{-\beta \omega}}\ 
\frac{\sin(\ell \omega)}{\ell \omega}\ ,
\end{eqnarray}
where $\ell$ denotes the modulus of the displacement vector $\vec\ell$;
then, recalling (\ref{19}), for the Hilbert transform one similarly
finds:
\begin{equation}
{\cal K}_{ij}^{(\alpha\beta)}(z)=
\delta_{ij}\ {\cal K}^{(\alpha\beta)}(z)\ ,\qquad
{\cal K}^{(\alpha\beta)}(z)=\frac{P}{\pi i}\int_{-\infty}^{\infty} dw\ 
\frac{ {\cal G}^{(\alpha\beta)}(w) }{w-z}\ .
\label{27}
\end{equation}
With these results and taking into account that 
$\sum_k \psi^{(\xi)}_{ki}\, \psi^{(\xi')}_{kj}=\psi^{(-\xi)}_{ij}\, \delta(\xi+\xi')$,
the Kossakowski matrix takes the more explicit form:
\begin{equation}
C_{ij}^{(\alpha\beta)}=
C^{(\alpha\beta)}_+\, \delta_{ij}-iC^{(\alpha\beta)}_-\, \sum_{k=0}^3\epsilon_{ijk}\, n_k 
+ \big[C^{(\alpha\beta)}_0-C^{(\alpha\beta)}_+]\, n_i\, n_j\ ,
\label{28}
\end{equation}
where
\begin{equation}
C^{(\alpha\beta)}_\pm=I^{(\alpha\beta)}_\pm\ \cos\big(\omega_{\alpha\beta}\,\Delta t/2\big)
+i\, I^{(\alpha\beta)}_\mp\ \sin\big(\omega_{\alpha\beta}\,\Delta t/2\big)\ ,\qquad
\omega_{\alpha\beta}\equiv\omega_\alpha-\omega_\beta\ ,
\label{29}
\end{equation}
with
\begin{equation}
I^{(\alpha\beta)}_\pm={\Delta t\over 4\pi}\int_{-\infty}^\infty d\omega\ 
\Big[{\cal G}^{(\alpha\beta)}(\omega)\pm{\cal G}^{(\alpha\beta)}(-\omega)\Big]\, 
{\sin\big[(\omega-\omega_\alpha)\Delta t/2\big]\over(\omega-\omega_\alpha)\Delta t/2}\
{\sin\big[(\omega-\omega_\beta)\Delta t/2\big]\over(\omega-\omega_\beta)\Delta t/2}\ ,
\label{30}
\end{equation}
while
\begin{equation}
C^{(\alpha\beta)}_0\equiv I^{(\alpha\beta)}_0={\Delta t\over 4\pi}\int_{-\infty}^\infty d\omega\ 
\Big[{\cal G}^{(\alpha\beta)}(\omega)+{\cal G}^{(\alpha\beta)}(-\omega)\Big]\, 
\Bigg[{\sin\big(\omega\Delta t/2\big)\over \omega\Delta t/2}\Bigg]^2\ .
\label{31}
\end{equation}
Only the following combinations
${\cal G}^{(\alpha\beta)}_\pm(\omega)\equiv{\cal G}^{(\alpha\beta)}(\omega)\pm{\cal G}^{(\alpha\beta)}(-\omega)$
actually occur in the previous integrals, and from the explicit expressions in (\ref{26}) one obtains:
\begin{eqnarray}
\label{32}
&&{\cal G}^{(11)}_+={\cal G}^{(22)}_+={\omega\over 2\pi}
\bigg[{1+e^{-\beta\omega}\over 1-e^{-\beta\omega}}\bigg]\ ,\hskip 1cm 
{\cal G}^{(12)}_+={\cal G}^{(21)}_+={\omega\over 2\pi}
\bigg[{1+e^{-\beta\omega}\over 1-e^{-\beta\omega}}\bigg]\ 
\frac{\sin(\omega\ell)}{\omega\ell}\ ,\hskip 1cm \\
\label{33}
&&{\cal G}^{(11)}_-={\cal G}^{(22)}_-={\omega\over 2\pi}\ ,\hskip 3cm
{\cal G}^{(12)}_-={\cal G}^{(21)}_-={\omega\over 2\pi}\ \frac{\sin(\omega\ell)}{\omega\ell}\ ;
\end{eqnarray}
they contain the dependence on the bath temperature $1/\beta$
and on the separation $\ell$ between the two atoms.
Because of the presence of the Boltzmann factors, the integrals $I^{(\alpha\beta)}_{\pm,0}$
in (\ref{30}), (\ref{31}) can not in general be expressed in terms of elementary functions.
However, in the case of a bath at high temperature ({\it i.e.} for small $\beta$), the square bracket in
(\ref{32}) behaves as $2/\beta\omega$ and the above integrals can be explicitly evaluated
(see the Appendix).
In the physical situation for which $\ell\leq \Delta t$,%
\footnote{This condition assures that the two atoms actually feel the
presence of the quantum fields; indeed, due to relativistic causality \cite{Haag}, the fields would
not be able to interact with the atoms in the time interval $\Delta t$
if they were to far apart.}
one finds:
\begin{eqnarray}
\nonumber
&&I^{(\alpha\beta)}_+={1\over\pi\beta\,\omega_{\alpha\beta}\, \Delta t}\Bigg\{
{\sin\big(\ell\omega_\alpha/2\big)\over \ell\omega_\alpha/2}\
\sin\bigg[\Big(\omega_\alpha\big(1-\ell/\Delta t\big)-\omega_\beta\Big){\Delta t\over2}\bigg]\\
\label{34}
&&\hskip 5cm +
{\sin\big(\ell\omega_\beta/2\big)\over \ell\omega_\beta/2}\
\sin\bigg[\Big(\omega_\alpha-\omega_\beta\big(1-\ell/\Delta t\big)\Big){\Delta t\over2}\bigg]\Bigg\}
\ ,\qquad\\
\label{35}
&&I^{(\alpha\beta)}_-={1\over\pi\ell\,\omega_{\alpha\beta}\, \Delta t}\
\sin\bigg[{\omega_{\alpha\beta}(\Delta t-\ell)\over2}\bigg]\ 
\sin\bigg[{\ell\big(\omega_\alpha+\omega_\beta\big)\over2}\bigg]\ ,\\
\label{36}
&&I^{(\alpha\beta)}_0={1\over4\pi\beta}\
\bigg(2-{\ell\over\Delta t}\bigg)\ .\phantom{\Bigg\vert^|}
\end{eqnarray}

Inserting these results back in (\ref{29}) and (\ref{31}), one finally obtains
the explicit expression for the Kossakowski matrix $C^{(\alpha\beta)}_{ij}$,
in the large temperature limit.%
\footnote{Since this is an approximated result,
positivity of the matrix is not a priori guaranteed and should
be formally imposed in order to preserve the properties of the exact expression (\ref{20}).
In particular, positivity of the two diagonal submatrices
$C^{(\alpha\alpha)}_{ij}$, requires $\beta\omega_\alpha/2\leq 1$,
which are satisfied by the requirement of $\beta$ small.}

Coming now to the Hamiltonian contribution to the master equation,
one sees that the effective Hamiltonian $H_{\rm eff}$ in (\ref{15}) can be split
into two parts, $H_{\rm eff}={\tilde H}_S+H_{\rm eff}^{(12)}$,
the first is just a renormalization of the starting system Hamiltonian, 
while the second one represents an environment induced
direct coupling term for the two atoms.
The term ${\tilde H}_S$ has the same form as the Hamiltonian in (\ref{1})
but with redefined frequencies
\begin{equation}
\tilde\omega_\alpha=\omega_\alpha-i{\Delta t\over 2\pi}
\int_{-\infty}^\infty d\omega\ 
\Big[{\cal K}^{(\alpha\alpha)}(\omega)-{\cal K}^{(\alpha\alpha)}(-\omega)\Big]\, 
\Bigg[{\sin\big[(\omega-\omega_\alpha)\Delta t/2\big]\over(\omega-\omega_\alpha)\Delta t/2}\Bigg]^2\ .
\label{37}
\end{equation}
Recalling the definition of ${\cal K}^{(\alpha\alpha)}(\omega)$
in (\ref{27}), one sees that it can be split as:
\begin{eqnarray}
\nonumber
{\cal K}^{(\alpha\alpha)}(\omega)&=&
\frac{1}{2\pi^2 i}\Bigg[ P
\int_0^\infty dz\
\frac{z}{z-\omega}\\
\label{38}
&&\hskip 2.5cm +P\int_0^\infty dz\ \frac{z}{1-e^{\beta z}}
\Bigg(\frac{1}{z+\omega}-\frac{1}{z-\omega}\Bigg)\Bigg]\ ,
\end{eqnarray}
into a vacuum and a temperature-dependent piece.
Although not expressible in terms of simple functions, the temperature
dependent second term is a finite, odd function of $\omega$; on the contrary,
the remaining, vacuum contribution in (\ref{38}) results divergent, and
therefore so are the shifted frequencies $\tilde\omega_\alpha$.
As a consequence, the definition of effective Hamiltonian
$H_{\rm eff}$ requires the introduction 
of a suitable cutoff and a renormalization procedure.
This is not a surprise: the appearance of the
divergences is due to the non-relativistic treatment of
the two-level atoms, while any sensible calculation
of energy shifts would have required the use 
of quantum field theory techniques \cite{Milonni}.
In order to make $H_{\rm eff}$ well defined we follow a simple prescription:
perform a suitable temperature independent subtraction, so that
the expressions in (\ref{37}) reproduce the correct
quantum field theory result, obtained by considering
the external fields in the vacuum state.

The induced two-atom interaction term $H^{(12)}_{\rm eff}$ can instead be expressed as
\begin{equation}
H^{(12)}_{\rm eff}=\sum_{i,j=1}^3 {\cal H}^{(12)}_{ij}\ \sigma_i^{(1)}\otimes\sigma_j^{(2)}\ ,
\label{39}
\end{equation}
where
\begin{equation}
{\cal H}^{(12)}_{ij}=\bigg(\cos\bigg[{\omega_{12}\Delta t\over2}\bigg]\ \delta_{ij}
+\sin \bigg[{\omega_{12}\Delta t\over2}\bigg]\, \sum_{k=0}^3\varepsilon_{ijk}\, n_k\bigg)J_+
+\bigg(J_0-\cos\bigg[{\omega_{12}\Delta t\over2}\bigg]\, J_+\bigg)\ n_i n_j\ ,
\label{40}
\end{equation}
with
\begin{eqnarray}
\label{41}
&&J_+=-i{\Delta t\over 4\pi}\int_{-\infty}^\infty d\omega\ 
\Big[{\cal K}^{(12)}(\omega)+{\cal K}^{(12)}(-\omega)\Big]\, 
{\sin\big[(\omega-\omega_1)\Delta t/2\big]\over(\omega-\omega_1)\Delta t/2}\
{\sin\big[(\omega-\omega_2)\Delta t/2\big]\over(\omega-\omega_2)\Delta t/2}\ ,\qquad\\
\label{42}
&&J_0=-i{\Delta t\over 8\pi}\int_{-\infty}^\infty d\omega\ 
\Big[{\cal K}^{(12)}(\omega)+{\cal K}^{(12)}(-\omega)\Big]\,
\bigg[{\sin\big(\omega\Delta t/2\big)\over \omega\Delta t/2}\bigg]^2\ .\phantom{\Bigg|^|}
\end{eqnarray}
Also ${\cal K}^{(12)}(\omega)$ can be split as in (\ref{38})
into a temperature dependent term, odd in $\omega$, and a vacuum piece. Clearly, only this second
contribution enters the above integrals $J_{+,0}$; it is finite (for non vanishing atom separation)
and with the help of (\ref{32}) can be explicitly computed:
\begin{equation}
{\cal K}^{(12)}(\omega)+{\cal K}^{(12)}(-\omega)=
\frac{P}{2\pi^2 i}\int_{-\infty}^{\infty} dz\ 
\frac{z}{z+\omega}\ {\sin\ell z\over \ell z}= {1\over 2\pi i}{\cos \ell\omega \over \ell}\ .
\label{43}
\end{equation}
Inserting this result in (\ref{41}) and (\ref{42}), one finally obtains, 
again for $\ell\leq\Delta t$ (see Appendix):
\begin{equation}
J_+=-{1\over 2\pi\ell\omega_{12}\Delta t}\ \cos\bigg[{\big(\omega_1+\omega_2\big)\ell\over 2}\bigg]\
\sin\bigg[{\omega_{12}\big(\Delta t-\ell\big)\over 2}\bigg]\ ,\qquad
J_0={1\over 8\pi}\bigg({1\over \Delta t}- {1\over \ell}\bigg)\ .
\label{44}
\end{equation}
We are now ready to discuss the entanglement properties of the time evolution generated by the master
equation (\ref{14})-(\ref{16}).

\section{Environment entanglement generation}

In order to study the entanglement power of the thermal bath made of free
quantum fields, we shall focus on the small $t$ behaviour of the dynamics generated by
the equation (\ref{14}); our aim is to investigate whether the two independent atoms
can get entangled by the action of the environment in which they are immersed
at the beginning of their dissipative evolution.
Then, without loss of generality, one can limit the considerations
to pure, separable initial states, and therefore take: 
\begin{equation}
\label{45}
\rho(0)=\vert \varphi\rangle\langle \varphi\vert\otimes 
\vert \psi\rangle \langle \psi\vert\ ,
\end{equation}
with $\vert \varphi\rangle$, $\vert \psi\rangle$ given single atom states;
indeed, if the environment is unable to create 
entanglement out of pure states, 
it will certainly not correlate their mixtures.

Since we are dealing with a couple of two-level systems, one can use
partial transposition as a criterion for entanglement creation \cite{Peres,Horodecki}.
More precisely, the environment is able to create quantum
correlations between the two atoms if and only if the operation
of partial transposition spoils the positivity of the
state $\rho(t)$.

The presence of negative eigenvalues in the partially
transposed reduced density matrix $\hat\rho(t)$ can be ascertained
by looking at the sign of the average
\begin{equation}
{\cal A}(t)=\langle\chi\vert\, \hat{\rho}(t)\, \vert\chi\rangle\ ,
\label{46}
\end{equation}
with $\vert\chi\rangle$ a four-dimensional vector. 
Indeed, choose $\vert\chi\rangle$ to be orthogonal to
$\vert \varphi\rangle\otimes\vert \psi\rangle$, so that the above average
 initially banishes, ${\cal A}(0)=\,0$.
Then, the two atoms, initially prepared in a 
state $\rho(0)\equiv\hat\rho(0)$ as in (\ref{45}), will surely become entangled
if $|\chi\rangle$ can be further chosen so that  $\partial_t {\cal A}(0)<0$.
From this condition, a simple test for entanglement creation involving
the elements of the Kossakowski matrix (\ref{28}) and of the effective
interaction Hamiltonian (\ref{39}) can then be extracted \cite{BFP,BLN}.
It explicitly reads:
\begin{equation}
\langle u | C^{(11)} | u \rangle \, \langle v | \big({C^{(22)}}\big)^T | v \rangle <
\big|\langle u | {\cal R}e\big(C^{(12)}+i H^{(12)}_{\rm eff} \big)| v \rangle \big|^2\ ,
\label{47}
\end{equation}
where $T$ means matrix transposition;
the three-dimensional vectors $|u\rangle$ and $|v\rangle$ 
contain the information about the starting factorized state (\ref{45});
in fact, their components can be expressed as:
\begin{equation}
u_i=\langle \varphi| \sigma_i |\varphi_\perp\rangle\ ,\quad
v_i=\langle \psi_\perp| \sigma_i |\psi\rangle\ ,
\label{48}
\end{equation}
where $|\varphi_\perp\rangle$ and $|\psi_\perp\rangle$ are the orthonormal
complement to the initial atom states $|\varphi\rangle$ and $|\psi\rangle$, respectively.
Therefore, the external quantum fields will be able to initially entangle the two atoms
evolving with the Markovian dynamics generated by (\ref{14}) if there exists an
initial state of the form (\ref{45}) for which the inequality (\ref{47})
is satisfied.

In order to obtain a manageable expression for it, we first note
that, without loss of generality,
the unit vector $\vec n$ that defines the
atom Hamiltonian in (\ref{1}) can be 
oriented along the third axis. Further, as initial atom state
we shall choose
$\rho(0)=|-\rangle\langle -|\otimes |+\rangle\langle +|$,
constructed out of the eigenstates of the single atom Hamiltonian,
$\sigma_3\, |\pm\rangle=\pm|\pm\rangle$. As a consequence,
recalling (\ref{48}), one finds that the three-dimensional vector $|u\rangle$
has components $u_i=\{1, -i, 0\}$, and further $v_i=u_i$.
Then, using the explicit expressions for the elements of
the Kossakowski matrix $C^{(\alpha\beta)}_{ij}$ and of the induced interaction Hamiltonian
$H^{(12)}_{\rm eff}$, the inequality (\ref{47}) reduces to
\begin{equation}
\bigg(1-{\beta\omega_1\over2}\bigg)\bigg(1+{\beta\omega_2\over2}\bigg) <
\pi^2\beta^2\Big[ \big(I^{(12)}_+\big)^2 + 4\big(J_+\big)^2 \Big]\ .
\label{49}
\end{equation}
Notice that the l.h.s. of this expression is positive, since as discussed
in the previous Section, complete positivity
requires $\beta\omega_\alpha/2\leq 1$.

As remarked at the end of Section 2, the parameter $\Delta t$ identifies
the time scale over which the presence of the environment is felt by the system
of the two atoms; clearly the weaker the coupling with the environment is, the
longer one needs to wait for the bath induced effects to become apparent.

Let us first discuss the standard {\it weak coupling limit} approximation;
in this case, one actually let the coupling constant
$\lambda$ to approach zero, so that changes in the two-atom density matrix
become visible only for infinitely large $\Delta t$. In this limit however,
the two integrals on the r.h.s. of (\ref{49}) become vanishingly small.%
\footnote{Indeed, in the limit $\Delta t \to\infty$ both integrals vanish
since, for $\omega_1\neq\omega_2$, the two functions 
${\sin[(\omega-\omega_1)\Delta t/2]/[(\omega-\omega_1)\Delta t/2]}$ and
${\sin[(\omega-\omega_2)\Delta t/2]/[(\omega-\omega_2)\Delta t/2]}$ have
disjoint supports.}
Thus, for atoms with unequal frequencies,
the inequality can never be satisfied, and thus no entanglement is generated.
On the contrary, when the two frequencies coincide, $\omega_1=\omega_2=\omega$,
the condition (\ref{49}) becomes:
\begin{equation}
1-\bigg({\beta\omega\over2}\bigg)^2 <
\bigg[ {\sin\big(\omega\ell\big)\over \omega\ell} \bigg]^2
+{\beta^2\over4}\, \bigg[ {\cos\big(\omega\ell\big)\over \ell} \bigg]^2\ .
\label{50}
\end{equation}
This result generalizes the one discussed in \cite{BF4}, where the contribution
of the environment induced interaction Hamiltonian
(the second term in the r.h.s. of (\ref{49})) was neglected. In particular, one sees that, in this case,
for any given (small) inverse temperature $\beta$, there
is always an atom separation $\ell$ below which the inequality (\ref{50}) is satisfied,
and therefore entanglement created between the two atoms. This phenomenon
is forbidden only for infinitely large separation or infinitely large temperature,
in which case the environment induced decoherence and noisy effects dominate.

The sharp dependence of the entanglement capability of the environment on
the atom frequencies in the {\it weak coupling limit} approach is however striking;
it originates in the elimination of fast oscillating terms in the reduced
two-atom dynamics through an ergodic average,
a procedure that is justified only in the limit of a vanishing $\lambda$ and very fast
decay correlations in the environment.

Instead, if the coupling of the atoms to the bath is weak, but not infinitesimally small,
environment induced changes in the atom density matrix $\rho(t)$ can be seen on
finite time intervals $\Delta t$. In this case, it is the full condition
(\ref{49}) that regulates the entanglement capability of the thermal bath.
One can check that indeed this inequality can be satisfied even for $\omega_1\neq\omega_2$,
and therefore that a bath made of thermal quantum fields can correlate two unequal atoms.

In order to show this, let us first note that the Hamiltonian contribution in (\ref{49}),
being positive, can only enhance entanglement generation; this is the result of the
hermiticity of the induced coupling term $H^{(12)}_{\rm eff}$ in (\ref{39}).%
\footnote{In general, the dissipative and Hamiltonian contributions in the r.h.s. of (\ref{47})
can destructively interfere making the inequality harder to be satisfied and thus
reducing the entanglement power of the environment.}
One can therefore limit the considerations to a simpler inequality, in which
the term $(J_+)^2$ is neglected; when this reduced condition is satisfied,
also the full one in (\ref{49}) will clearly be.
Recalling (\ref{34}), and keeping for simplicity only first order terms in $\ell$, 
the condition for environment assisted entanglement generation reduces to:
\begin{equation}
\bigg(1-{\beta\omega_1\over2}\bigg)\bigg(1+{\beta\omega_2\over2}\bigg) <
\bigg[{\sin\big(\omega_{12}\,\Delta t/2\big) \over \omega_{12}\,\Delta t/2}\bigg]^2
-\bigg({\ell\over\Delta t}\bigg)\,
{\sin\big(\omega_{12}\,\Delta t\big) \over \omega_{12}\,\Delta t}\ .
\label{51}
\end{equation}
For given bath temperature and atom frequencies, this condition is satisfied
provided a sufficiently small time interval $\Delta t$ results from the
coupling with the environment. Further, the smaller the atom separation is,
the easier the condition of (\ref{51}) will be met. 

Finally, note that, in contrast to situation encountered in 
the {\it weak coupling limit} approximation, here there is no sharp
change between the regime of entanglement generation and the one of solely decoherence;
the transition is smoothly regulated by the coarse graining parameter $\Delta t$,
{\it i.e.} ultimately by the strength of the coupling of the atoms to the environment.

\section{Discussion}

We have seen that two atoms, 
prepared initially in a separable state,
can get entangled as a result of their independent interaction with a common bath made
of thermal quantum fields even when their internal frequencies are unequal.
This result is based on a novel Markovian approximation of the reduced atom dynamics,
that allows an explicit dependence on the time scale $\Delta t$, measuring the interval over which 
the atoms feel the presence of the environment. 

This conclusion contrasts with the one obtained through
the usual {\it weak coupling limit} approach to the
atom reduced dynamics; in that case, the entanglement power of the
external environment is reduced to zero for atoms with unequal frequencies
as a consequence of the procedure of taking the ergodic average. 
In the light of the results presented in the previous Section, 
this conclusion appears however a mathematical artifact,
originating in letting
$\lambda$ to zero and $\Delta t$ to infinity,%
\footnote{Because of the Riemann-Lebegue lemma,
the ergodic average, on which the so-called ``rotating wave approximation''
is based, is strictly justified only in the limit $\lambda\to 0$ and $\Delta t\to \infty$.}
conditions hardly met in actual physical situations.
Instead, for weakly coupled baths with finite $\Delta t$,
environment assisted entanglement generation is always allowed, and can be controlled
through the external parameters, the bath inverse temperature $\beta$ and
the atom spatial separation $\ell$.

In the high temperature case ({\it i.e.} $\beta$ small) and arbitrary $\Delta t$,
we have explicitly shown that this conclusion holds because of the condition (\ref{51}). 
Similarly, in situations allowing a large, but finite $\Delta t$,
a different approximation of the full entanglement condition (\ref{47}) 
can be given; it can be obtained using techniques and procedures analogous to the ones
discussed in the previous Sections. Neglecting again the Hamiltonian contribution, one explicitly finds:
\begin{equation}
\Big(1-R_1\Big)\Big(1+R_2\Big) < {1\over4}
\bigg[{\sin\big(\omega_{12}\,\Delta t/2\big) \over \omega_{12}\,\Delta t/2}\bigg]^2\
\bigg[ \bigg( {\omega_1\over\omega_2}\, {R_2\over R_1}\bigg)^{1/2}\, S_1
+\bigg( {\omega_2\over\omega_1}\, {R_1\over R_2}\bigg)^{1/2}\, S_2\bigg]^2\ ,
\label{52}
\end{equation}
where
\begin{equation}
R_\alpha= {1-e^{-\beta\omega_\alpha} \over 1+e^{-\beta\omega_\alpha}}\ , \qquad 
S_\alpha= {\sin\big(\omega_\alpha\ell\big) \over \omega_\alpha\ell}\ ,
\qquad \alpha=1,2\ .
\label{53}
\end{equation}
It is a further generalization of the condition discussed in \cite{BF4} in the case
of identical atoms, to which it reduces for $\omega_1=\omega_2$ and 
$\Delta t$ infinite.
Although valid only for large (but finite) $\Delta t$, 
it can always be satisfied with suitably chosen $\beta$ and $\ell$.
In particular, (\ref{52}) is always true in the zero temperature case, {\it i.e.} in the limit
$\beta\to\infty$; 
in other words, a bath made of quantum fields in the vacuum state is always able to
generate entanglement, for any finite spatial separation of the two atoms.

All the above considerations are based on the condition
(\ref{47}) for entanglement enhancement; when satisfied, it assures that
quantum correlations among the two atoms are generated as soon as $t>0$.
It is however unable to determine the fate of this quantum correlations
as time increases and in particular in the asymptotically long time regime.
On general grounds, one expects that the effects 
of decoherence and dissipation that
counteract entanglement production
be dominant at large times, so that 
no entanglement is left in the end.
There are however instances in which the entanglement generated at the beginning 
of the evolution persists also for
asymptotically long times \cite{BF5,BF6,BF}. In order to fully clarify this situation,
a complete study and classification of the set of the equilibrium states of 
the refined master equation (\ref{10})-(\ref{13}) is necessary.%
\footnote{Only partial results on the classification the equilibrium states
of completely positive quantum dynamical semigroups have been so far obtained \cite{Spohn,Frigerio}.}
Work on this topic is presently in progress and will be reported elsewhere.

\vskip 2cm
\section*{Appendix}

We indicate here how to compute the integrals that appear in the expressions
of the Kossakowski matrix $C_{ij}^{(\alpha\beta)}$, (\ref{30}) and (\ref{31}), and in that of the effective
Hamiltonian interaction term $H_{\rm eff}^{(12)}$, (\ref{41}) and (\ref{42}).
In the high temperature case, the explicit evaluation of (\ref{30}) involves the
computation of integrals of the following two types:
\begin{eqnarray}
\label{54}
&&I_1=\int_{-\infty}^{+\infty} dx\ \sin(c\, x)\ {\sin (x-a)\over x-a}\ {\sin(x-b)\over x-b}\ ,\\
\label{55}
&&I_2=\int_{-\infty}^{+\infty} dx\ {\sin(c\, x)\over x}\ {\sin (x-a)\over x-a}\ {\sin(x-b)\over x-b}\ ,
\end{eqnarray}
with $a$, $b$, $c$ positive constants. By decomposing the products of
trigonometric functions in terms of linear combinations of sines and cosines, 
one can split {\it e.g.} $I_1$ into the sum of three simpler integrals:
\begin{equation}
I_1=I_0+I(c)-I(-c)\ ,
\label{56}
\end{equation}
with
\begin{equation}
I_0={\cos(a-b)\over 2}\int_{-\infty}^{+\infty} dx\ {\sin(c\, x)\over (x-a)(x-b)}\ ,\qquad
I(c)={1\over4}\int_{-\infty}^{+\infty} dx\ {\sin\big[(2-c)x-a-b\big]\over (x-a)(x-b)}\ .
\label{57}
\end{equation}
By first changing the integration variable to $y=(2-c)x-a-b$ in $I(c)$, with $c\leq2$, and then reducing
the denominators in partial fractions in both integrands, one can express
$I_0$ and $I(c)$ as combinations of the following integral ({\it e.g.} see \cite{Prudnikov}):
\begin{equation}
\int_{-\infty}^{+\infty} dx\ {\sin(\alpha\, x)\over x+z}=\pi\, \cos(\alpha z)\ ,\quad \alpha>0\ .
\label{58}
\end{equation}
Explicitly, one finds:
\begin{eqnarray}
\label{59}
&&I_0=-\pi{\cos(a-b)\over a-b}\ \sin\big[(a+b)c/2\big]\,\sin\big[(a-b)c/2\big] \ ,\\
\label{60}
&&I(c)={\pi\over 4(a-b)}\Big( \cos\big[a(c-1)+b\big]-\cos\big[a+b(c-1)\big]\Big)\ ,
\end{eqnarray}
so that, recalling (\ref{56}), one finally obtains:
\begin{equation}
I_1=\pi \sin\bigg[{(a+b)c\over2}\bigg]\,
{\sin\big[(a-b)(1-c/2)\big]\over (a-b)}\ .
\label{61}
\end{equation}
This result holds for $c\leq2$; when $c>2$, one is forced to use a different integration variable
in the expression of $I(c)$ in (\ref{57}), $y'=(c-2)x+a+b$, and as a result ends up with
a vanishing value for $I_1$. As a function of the parameter $c$,
the integral $I_1$ is however continuous, since the expression in (\ref{61}) also vanishes
at the boundary point $c=2$.

From the result (\ref{61}), one further obtains:
\begin{equation}
\lim_{c\to 0} \bigg({I_1\over c}\bigg)=\int_{-\infty}^{+\infty} dx\ x\ {\sin (x-a)\over x-a}\
{\sin(x-b)\over x-b}
=\pi \bigg({a+b\over 2}\bigg)\, {\sin(a-b)\over a-b}\ .
\label{62}
\end{equation}

The integral $I_2$ in (\ref{55}) can be evaluated using similar manipulations. 
When $c\leq2$, one explicitly finds
\begin{equation}
I_2={\pi\over a-b}\bigg( {\sin(ac/2)\over a}\, \sin\big[a(1-c/2)-b\big] +
{\sin(bc/2)\over b}\, \sin\big[a-b(1-c/2)\big]\bigg)\ ,
\label{63}
\end{equation}
while for $c>2$, a simpler expression holds:
\begin{equation}
I_2=\pi\ {\sin a\over a}\, {\sin b\over b}\ .
\label{64}
\end{equation}
Here again one sees that $I_2$ is a continuous function of $c$, since the expression
in (\ref{63}) reduces to the one in (\ref{64}) at the boundary value $c=2$.
Further, from the expression in (\ref{63}), one easily obtains the following
limiting results:
\begin{equation}
\lim_{c\to 0} \bigg({I_2\over c}\bigg)=\int_{-\infty}^{+\infty} dx\ {\sin (x-a)\over x-a}\ {\sin(x-b)\over x-b}
=\pi \, {\sin(a-b)\over a-b}\ .
\label{65}
\end{equation}
and similarly,
\begin{equation}
\lim_{a,b\to 0} I_2=\int_{-\infty}^{+\infty} dx\ {\sin(c\, x)\over x}\bigg({\sin x\over x}\bigg)^2
=\pi \,c\bigg(1-{c\over 4}\bigg)\ .
\label{66}
\end{equation}

The integrals appearing in the evaluation of the Hamiltonian contribution $H_{\rm eff}^{(12)}$
can instead be all reduced to expressions of the form:
\begin{equation}
J=\int_{-\infty}^{+\infty} dx\ \cos(c\, x)\ {\sin (x-a)\over x-a}\ {\sin(x-b)\over x-b}\ .
\label{67}
\end{equation}
With the help of manipulations similar to the one used above, $J$ can be reduced to combinations
of the following integral \cite{Prudnikov}:
\begin{equation}
\int_{-\infty}^{+\infty} dx\ {\cos(\alpha\, x)\over x+z}=\pi\, \sin(\alpha z)\ ,\quad \alpha>0\ .
\label{68}
\end{equation}
When $c\leq 2$, the integral in (\ref{67}) can be cast in the following form:
\begin{equation}
J=\pi \cos\bigg[{(a+b)c\over2}\bigg]\,
{\sin\big[(a-b)(1-c/2)\big]\over (a-b)}\ ,
\label{69}
\end{equation}
while it vanishes for $c>2$. In the limit of vanishing $a$ and $b$, it reduces to
$J=\pi(1-c/2)$.

\vfill\eject

\end{document}